# Mechanism of all-optical control of ferromagnetic multilayers with circularly polarized light

R. Medapalli,[1,*] D. Afanasiev,[2] D. K. Kim,[1] Y. Quessab,[1, 3] S. Manna,[1] S. A. Montoya,[1] A. Kirilyuk,[2] Th. Rasing,[2] A. V. Kimel,[2] and E. E. Fullerton[1]

[1]*Center for Memory and Recording Research, University of California, La Jolla, USA*
[2]*Radboud University Nijmegen, Institute for Molecules and Materials, Heyendaalseweg 135, Nijmegen, The Netherlands*
[3]*Institut Jean Lamour, UMR CNRS 7198, Universite de Lorraine, Vandoeuvre-l_es-Nancy, France*
(Dated: July 6, 2016)

Time-resolved imaging reveals that the helicity dependent all-optical switching (HD-AOS) of Co/Pt ferromagnetic multilayers proceeds by two stages. First one involves the helicity independent and stochastic nucleation of reversed magnetic domains. At the second stage circularly polarized light breaks the degeneracy between the magnetic domains and promotes the preferred direction of domain wall (DW) motion. The growth of the reversed domain from the nucleation cite, for a particular helicity, leads to full magnetic reversal. This study demonstrates a novel mechanism of HD-AOS mediated by the deterministic displacement of DWs.

Controlling the magnetic state of media with ultrashort laser pulses is a rapidly growing research area which promises to revolutionize information processing by achieving the fastest possible and least dissipative magnetic recording[1,2]. The interaction of light with electrons, within the electric-dipole approximation, conserves the electron spin. As a result, the mechanisms allowing femtosecond optical control of magnetism are among the most heavily debated fundamental questions in contemporary condensed matter research. The seminal experimental observation of a subpicosecond laser-induced demagnetization[3] in metals showed the fastest way of quenching the total spin angular momentum. This study triggered scientific exploration into controlling spins with ultrashort laser pulses and led to many intriguing experimental observations[4-6], including the discovery of the helicity dependent all-optical magnetic switching (HD-AOS) in ferrimagnetic GdFeCo films[7]. It was demonstrated that a single laser pulse brings this ferrimagnet into a strongly non-equilibrium[8] state by full demagnetization of just one of the spin-sublattices[9]. Relaxation from this state is accompanied by the magnetization reversal of the two magnetic sub-lattices[10, 11].

This explanation, however, implies that the materials exhibiting HD-AOS must have at least two non-equivalent and antiferromagnetically coupled magnetic sublattices[11]. Nevertheless, recently it was reported that in ferromagnetic Co/Pt the magnetization can be also reversed upon excitation with a sequence of circularly polarized femtosecond laser pulses, where the final state is deterministically determined by the helicity of the laser pulses [12, 13]. The mechanism developed earlier for ferrimagnets, based on the antiferromagnetic exchange between two sublattices, is not applicable in this case. To date, deterministic optical switching of Co/Pt films has only been demonstrated for multiple optical pulses[12,13] while neither optical control by a single pulse nor the dynamics of the deterministic magnetization reversal as seen in GdFeCo films, have been reported so far. Here we report the dynamics of the magnetization reversal in a Co/Pt multilayer, starting from the ultrafast magnetic response to a single pump pulse excitation to the formation of equilibrium reversed magnetic domain.

The sample studied is a multilayer stack of Ta(5 nm)/Pt(5 nm)/[Co(0.4 nm)/Pt(0.7 nm)]$_{x3}$/Pt(2 nm), grown by magnetron sputtering on a 500 µm thick soda lime glass substrate. The film has strong perpendicular magnetic anisotropy, so the magnetization easy axis is normal to the film and two stable magnetizations orient along or against the sample normal: $M_-$ or $M_+$, respectively. The sample was excited by either left-handed ($\sigma_-$) or right-handed $\sigma_+$) circularly polarized laser pulses with a central wavelength of 800 nm. In order to study the sub-nanosecond dynamics triggered by a single pump pulse, a time-resolved single-shot magneto-optical imaging technique[8] was used. Linearly polarized laser pulses with a duration of 4.0 ps were used to probe the laser-induced picosecond magnetization dynamics. An external magnetic field of ±800 Oe, larger than coercivity of the sample[14], was applied to saturate the sample after each pulse and was removed before every pump-probe event was measured. To reveal millisecond magnetization dynamics an halogen lamp was used as a source of probe light. The magnetic changes in the pump area were traced as a function of exposure time with a resolution close to 90 ms. A CCD camera in combination with collective optics and two nearly crossed polarizers was used to image magnetic domains in the sample via the magneto-optical Faraday effect with a resolution of 1 µm.

First, we experimentally determined the optimal conditions for HD-AOS. For this we choose a combination of right-circularly polarized pump beam and a magnetic domain with magnetization $M_-$, denoted as ($\sigma_+$, $M_-$). By slowly sweeping the beam at a repetition rate of 500 Hz across the magnetized area we obtained different magnetic domain patterns (see Fig. 1a) which depend on the fluence and duration of the pump pulse. The patterns represent a multi-domain magnetic state with different fractions of domains having either initial or reversed magnetizations. To characterize the photo-induced state, images were digitized and integrals were taken over the areas where the magneto-optical contrast had been changed. The result of the integration, that represents the average net magnetization <M>, is considered as a measure of the efficiency of the switching in this paper. The case of <M> = 0 corresponds to a multidomain state with zero average magnetization. The homogeneously magnetized domains with non-reversed or reversed magnetization correspond to <M> = -1 and <M> = +1, respectively. From Fig.1b, it is clearly seen that, unlike previous experiments on ferrimagnetic GdFeCo alloys[15], the effciency of the switching in the ferromagnetic bilayers increases if the pump duration increases[16], with a saturation above 3.5 ps. Figure 1c shows that there is also a strong fluence dependence. To summarize the experimental observations we plot a phase diagram (see Fig.1d) of the switching efficiency <M>. It is seen that efficient switching (<M> ≥ 0.63) is observed only in a narrow range of pump fluences and for a pulse length larger than 1.6 ps. For none of the fluences and pulse durations helicity-independent switching was observed as seen in GdFeCo films.

To determine the effect of a single pump pulse for an optimal combination of duration (4.0 ps) and fluence (0.5 mJ/cm$^2$), we performed single-shot experiments. Figure 2a shows the images both before and after the laser excitation. The area exposed to the pump pulse is shown by a dashed circle. No apparent difference for images before and after the single pump pulse is seen. However, it is clearly visible that after an action of 100 pulses a domain with the reversed magnetization is produced. To further reveal if any helicity-dependent dynamics is present in the sub-ns time-domain, a time-resolved single-shot imaging experiment, for two opposite helicities ($\sigma_-$, $\sigma_+$) of the pumping light and two initial states of the magnetization, ($M_-$, $M_+$) was performed. The magnetization dynamics for each of the combinations of pump helicity and initial magnetization is plotted in Fig. 2b. One can clearly see that the laser excitation results in a rapid partial demagnetization (within 8 ps), followed by a slow recovery while no magnetization reversal is observed. Interestingly, changing the helicity of the pump light does not result in any pronounced difference in the photo-induced dynamics. The asymmetry in the signals for two opposite values of the magnetization originates from the intrinsically nonlinear response of the cross-polarized magneto-optical scheme[17]. To eliminate this and to obtain quantitative values of the ultrafast demagnetization, the difference was taken and the normalized value is plotted in Fig. 2c. It is seen that the ultrafast demagnetization reaches values up to 75%. The subsequent magnetization recovery takes more than 1 ns and demonstrates a double exponential relaxation, similar to that observed in Ref.[18]. Hence, these experiments emphasize the importance of multiple pulse excitation for HD-AOS in Co/Pt.

To explore how the repetition rate affects the switching, the sample was exposed to a sequence of ultrashort laser pulses of different repetition rates in therange 0.1 Hz to 1 kHz, while the conditions of the beam helicity and the magnetization ($\sigma_-$, $M_+$) were fixed. We find that exposing the sample to many pulses finally results in full magnetization reversal, even if pulses are separated by 10 s. Any effects of a pump-to-pump heat accumulation can be excluded on such a long time scale, since the metallic film has enough time to relax back to its initial state The photo-induced changes in the region exposed to the repetitive action of the laser pulses at the repetition rate of 1 Hz for various time intervals are shown in Fig. 3a. It is seen that the first visible changes in the magnetic contrast appear only after 60 s, i.e. when the sample has already been excited by 60 pulses. It is known that magnetization reversal is a first-order phase transition [19]. In the kinetics of such phase transitions one distinguishes two regimes: nucleation of the domains of the new phase and their subsequent growth. To study these two regimes we extracted the magnetic profile for a vertical line cross-section through the center of the pump exposed region. The line scans obtained from the images made at different times were merged together such that the resulting image shows how the domains with the reversed magnetization evolve in time. The evolutions of the domains under excitation by a sequence of laser pulses with different repetition rates are shown in Fig. 3b. In all the cases the multi-pulse dynamics is described by nucleation followed by a gradual growth of the reversed domain. To show the helicity dependence during the growth process, similar results for four different combinations of ($\sigma$, $M$) are shown in Fig. 2 of Ref.[14] for the case of 1 kHz. The number of pulses required to nucleate a reversed domain as well as the number of nuclei varied between experiments (see Fig.3b). For the case where multiple domains nucleate, they initially independently grow with subsequent pulses before merging into a single domain. Note that nucleation does not start immediately and requires a finite time before a very small reversed domain emerges. This time does not show any pronounced dependence on the repetition rate and illumination time and seems to be intrinsically stochastic.

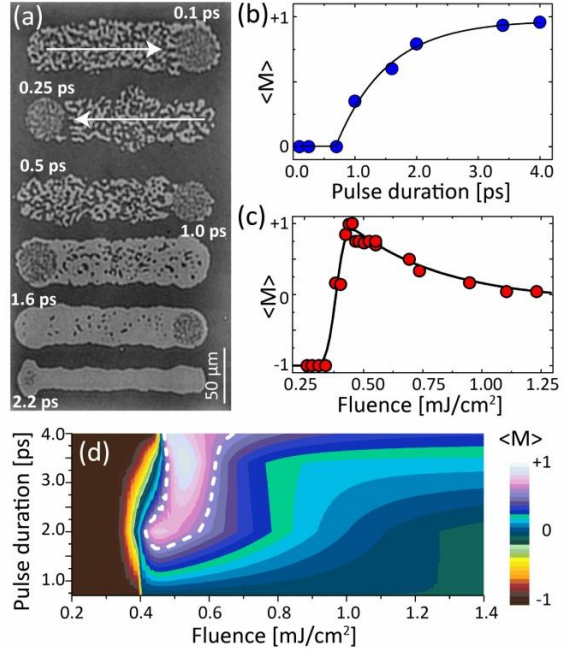

FIG. 1: (a) Magneto-optical image of the multi domain patterns induced by slow sweeping of the circularly polarized pulsed laser beam along the sample surface. Each pattern corresponds to a specific duration of the pump pulse. The pump repetition rate is 500 Hz, fluence is 0.5 mJ/cm$^2$. Arrow indicates direction of the sweep. (b) Average value of the photo-induced magnetization <M> as a function of the pulse duration. The pump fluence is 0.5 mJ/cm$^2$. (c) Average value of the photo-induced magnetization <M> as a function of the pump fluence. The duration of the pump pulse is 2.2 ps. (d) The phase diagram of the value <M> as a function of the pump pulse duration and fluence. The color bar indicates the value of <M>. The dashed contour shows the pump fluences and the pump pulse durations for which efficient switching was observed.

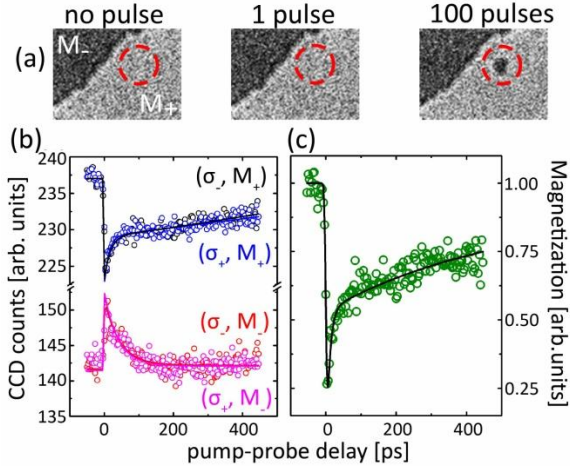

FIG. 2: (a) Magneto-optical images of the Co/Pt bilayer after illumination with a fixed number of circularly polarized ($\sigma_-$) pump pulses in the $M_+$-region. The area exposed to the pump pulses is shown by the dashed circle. The pump fluence is 0.5 mJ/cm$^2$, the pulse duration is 4.0 ps. (b) The single-shot dynamics of the average value of the magneto-optical contrast in the area affected by the pump pulse for different combinations of the light helicity ($\sigma_-$, $\sigma_+$) and magnetic ground state ($M_-$, $M_+$). The solid lines are guides to the eye. (c) The dynamics of the absolute demagnetization.

Figure 3c shows the typical dynamics of the <M> value, measured along the vertical cross-section. It is seen that the dynamics reveals two distinct regimes, each of which is well described by an exponential function. The first, fast, one with characteristic time $\tau_1$ describes the primary stage in the formation of the reversed magnetization. The second one, much slower, describes the domain growth and is characterized by the time constant $\tau_2$. The characteristic times $\tau_1$ and $\tau_2$ as a function of repetition period ($P$) are plotted in Fig. 3d. Both plots show a nearly linear dependence. This indicates that the repetition rate does not influence the multipulse dynamics. The slope of the line fit, in turn, reveals the number of the pulses necessary to accomplish the changes in magnetization in each of the regimes. The displacement of the reversed domain, within a characteristic time $\tau_1$ or $\tau_2$, along the line cross-section shown in Fig. 3a can be measured. The ratio of this (radial) displacement to the total number of pump pulses exposed to achieve this, yields the laser-induced displacement per pulse. Our experiments show that within the fast growth regime each laser pulse enlarges the size of the domain by 50±23 nm radially, acquiring a size of around 15 μm in diameter. During the final stage the enlargement caused by a single pulse is reduced down to 3.1±1.1 nm. The nucleation process is stochastic and independent of the helicity of the light, indicating that nucleation results from thermal-induced demagnetization. The nature of the initial nuclei is beyond the resolution of our optical system and remains unclear. These could be mesoscopic structures such as nanosized magnetic bubbles[20] or even magnetic skyrmions[21, 22]. Naturally, such nuclei are separated by an intermediate spin arrangement (e.g. domain wall) from the surrounding magnetization.

Contrary to the nucleation process, our experiments show that, once the nuclei are formed the subsequent domain growth does depend on the helicity of the light and is governed by the number of pump excitations. This is consistent with a field driven reversal in ultrathin ferromagnetic multilayers, where the growth of the reversed domains is determined by domain wall propagation rather than nucleation of small adjacent reversed domains[23]. In our optical switching experiments the growth process could be attributed to the cumulative displacement of domain walls induced by individual pump pulses and it is this helicity-dependent domain growth process which determines the final magnetic state of the HD-AOS process.

The implication of the above results is that the propagation direction of the domain due to optical excitation is helicity dependent. To directly demonstrate this we prepared a sample with a domain wall separating large $M_+$, and $M_-$-regions. We then fixed the center of the laser beam on the domain wall and recorded the photo-induced changes both on the helicity of the light and the number of pulses. The fluence of the pump pulses was fixed to a low value (0.4 mJ/cm$^2$) such that no significant nucleation events were observed at any stage of optical excitation. Hence, the changes in the magnetizations in the laser exposed region can safely be attributed to domain wall displacement. Figure 4a shows the images of the $M_+$, and $M_-$-regions, separated by the domain wall, taken before and after exposure to a sequence of 100, 600, and 3,600 $\sigma_-$-pulses. The dashed curve shows the initial position of the wall. The displacement of the domain wall, towards the $M_+$-region, is clearly visible and continues to move with increasing number of excitations. Figure 4b shows a similar result, for the case of $\sigma_+$-pulses, with the domain wall displacing in the opposite direction i.e. towards the $M_-$-region. The calculated average displacements per pulse, for either of the $\sigma_-$ or $\sigma_+$ cases, after pumping with 3,600 pulses is around 2 nm.

The direction of domain wall displacement is therefore directly linked to the helicity of the excitation light and therefore provides insight into the fundamental mechanism of HD-AOS. It is known that the action of a circularly polarized laser pulse on spins can be considered as a pulse of effective magnetic field, the inverse Faraday effect (IFE), the direction of which is determined by the helicity of the light[24]. The duration of the light-induced IFE field is given by the optical coherence time in the medium and in metals comparable to the pulse duration[15]. This opto-magnetic field can act on the magnetization in the domains and drive the motion of the domain walls, which would result in the enlargement of a domain with specific direction of the magnetization. In our experiments the duration of the optical pulse was set to 4.0 ps, which corresponds to a typical speed of domain walls close to 1km/s. The velocity is much larger than the typical velocity in ferromagnetic metallic thin films[25] close to the Walker limit (~100 m/s). This value is comparable to that recently observed in synthetic antiferromagnets[26] driven by current pulses. However, we do note that the width of a domain wall is on the order of the displacement per pulse.

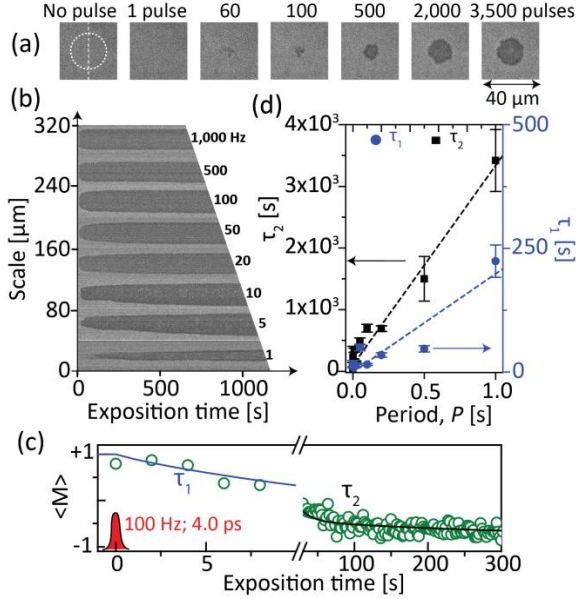

FIG. 3: (a) Magneto-optical images of laser-induced magnetic domains subjected to the repetitive action of the pump pulses ($\sigma_-$, $M_+$). The repetition rate is 1 Hz. The pump fluence is 0.5 mJ/cm$^2$, and the pulse duration is 4.0 ps. (b) The magnetization dynamics in a cross-section of the laser-excited area for different repetition rates. (c) Multipulse dynamics of the average magnetization <M> for the repetition rate 100 Hz. (d) The characteristic growth times $\tau_1$ and $\tau_2$ as a function of repetition period, $P$. The (dashed) straight lines are linear fits to the data.

Recently several microscopical mechanisms with thermal origin for domain wall motion (see Ref. [13] and references within), have been suggested. These mechanisms are based on the existence of thermal gradients between areas with oppositely oriented magnetizations. Due to magnetic circular dichroism, there will be differential absorption[27] of given circularly polarized pulses by the $M_-$ and $M_+$-regions. This differential absorption, on the order of 1.3 %, will result in slight temperature differences of opposite domains but will lead to a significant thermal gradient over the distance of the domain wall. This thermal gradient can potentially drive the domain wall locally from hotter to colder regions, which can explain the helicity dependent growth of the reversed domains. Finally, we note that we cannot rule out the direct interaction of the electric field of light with the spins via opto-magnetic effects[1, 24] causing the displacement of the walls at the order of the wall width. We also note that the recent modeling suggests that due to a strong spin-orbit coupling in Co/Pt multilayers, a single 0.1 ps pulse can reverse magnetization[28]. We do not find confirmation of this mechanism in our experiments.

In summary, we experimentally demonstrated the processes of magnetization reversal in a ferromagnetic multilayer by means of ultrashort laser pulses as external stimulus. For this, first we experimentally obtained the optimal pulse conditions for HD-AOS. Using these parameters, we determined the dynamics of the most efficient HD-AOS in Co/Pt multilayers. We clearly demonstrate that the observed effect is multipulse in nature, although it is independent of the multipulse accumulation of heat. It is shown that the magnetic switching proceeds via thermally-induced stochastic nucleation of reversed domains followed by a helicity-dependent deterministic growth. The controllability over the direction of the domain wall displacement from the nucleation sites is the key of the growth process. It consists of two distinct regimes and within each of them, the single pump pulse results in a domain wall's displacement by 50 and 3 nm, respectively. The controlled displacement per pulses in the faster growth regime are comparable to those achieved in external magnetic fields strengths near the Walker limit in ferromagnets or at the injection of spin-polarized current pulses into ferromagnetic nanowires.

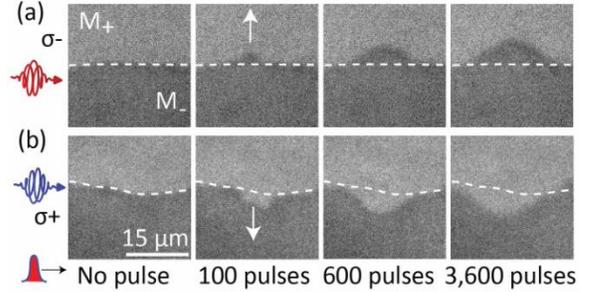

FIG. 4: (a) Magneto-optical images of laser-induced domainwall displacement at the region separating $M_+$ and $M_-$-domains, when subjected to the repetitive action of the $\sigma_-$-pump pulses. The repetition rate is 1 Hz. The pump fluence is 0.4 mJ/cm$^2$, the pulse duration is 4.0 ps. (b) Same as (a) but for the case of $\sigma_+$ pump pulse pulses. The dashed straight line is drawn to show the initial position of the domain wall. Note that the direction of displacement of the domain wall is controlled with the helicity of the laser pulses.

The authors would like to thank Prof. B.A. Ivanov for inspiring discussions and R. Descoteaux (CMRR UCSD), T. Toonen, S. Semin and A. van Etteger (Radboud University Nijmegen) for the technical support. This work was partially supported by NRI, the Offce of Naval Research MURI Program, the EU Seventh Framework Program grant agreement No. 281043 (FEMTOSPIN), the European Research Council grant agreement 339813 (Exchange) and 257280 (Femtomagnetism), and the FOM programmes SPIN and Exciting Exchange.

————————————

*rmedapalli@ucsd.edu